\documentstyle[12pt]{article}
\def\NPB{{Nucl. Phys.} {\bf B}}
\def\PLB{{Phys. Lett.}  {\bf B}}
\def\ZPC{{Z. Phys.} {\bf C}}
\def\be{\begin{equation}}
\def\ee{\end{equation}}
\def\bea{\begin{eqnarray}}
\def\eea{\end{eqnarray}}

\begin{document}

\rightline{DESY 97-150}
\rightline{August 1997}
\vspace{1.5cm}

\begin{center}
{\large \bf ON THE ODDERON INTERCEPT IN QCD}\footnote{Talk given by N.
Armesto at the {\it Madrid Workshop on low $x$ Physics} (Miraflores de
la Sierra, Spain, June 18th-21st 1997).}
\vspace{1cm}

N. Armesto\\
{\it II. Institut f\"ur Theoretische Physik, Universit\"at
Hamburg,\\
Luruper Chaussee 149, D-22761 Hamburg, Germany}\\
\vspace{0.5cm}
and\\
\vspace{0.5cm}
M. A. Braun\\
{\it Department of High Energy Physics, University of St.
Petersburg,\\ 198904 St. Petersburg, Russia}
\end{center}
\vspace{1.5cm}

The odderon singularity
is studied in perturbative QCD in the
framework of the Bartels-Kwieci\'nski-Praszalowicz (BKP) equation.
Arguments for the odderon intercept
being exactly equal to
unity are given. Besides, a variational method based on a
complete system of one-gluon functions is presented.
For the odderon, the highest
intercept
calculated by this method is $1-(N_c\alpha_{s}/\pi)\,0.45$. Comparison
to other calculations is shown.
\newpage

\section{Introduction}\label{sec:intro}

Regge theory \cite{regge} has been used during the last 30 years to
describe strong interaction at high energies and low transferred
momenta. The amplitude for the reaction $p_a p_b \longrightarrow
p_a^\prime p_b^\prime$, in the limit $s\gg m^2 \simeq -t$,
can be expressed as a sum over Regge trajectories
$j(t)=1+\omega(t)$ exchanged in the $t$-channel:
\be
A(s,t)=\sum_{j; p=\pm} \xi_{j_p(t)}^p\ s^{j_p(t)}\ g_1^p(t)g_2^p(t),
\label{eq:1}
\ee
with $s=(p_a+p_b)^2$ and $t=(p_a-p_a^\prime)^2$ the Mandelstam
variables and $\xi_{j}^p=i-(\cos{\pi j} +p)/\sin{\pi j}$ the signature
factor.

The amplitude can be decomposed as a sum over parts with definite
signature, i.e. definite
behaviour under the exchange $s \longrightarrow -s$:
\be
A(s,t)=A^+(s,t)+A^-(s,t), \ \ A^\pm(s,t) =\pm A^\pm(-s,t).
\label{eq:2}
\ee

As the total cross section can be related to the amplitude via the
optical theorem, $\sigma_{tot} \propto \mbox{Im}\ A(s,0)/s$, the
behaviour at high energies will be determined by those Regge
trajectories exchanging vacuum quantum numbers and
with highest intercept $\omega(0) +1= j(0)$. For positive signature
(contributing equally both to $pp$ and $\overline{p}p$ scattering) the
trajectory with the highest intercept is called the pomeron and its
intercept has been determined \cite{pom} to be supercritical (i.e. $>1$):
$\omega(t)\simeq 0.08+ (0.25\
\mbox{GeV}^{-2})\ t$. The trajectory with negative signature
(contributing to the difference between $pp$ and $\overline{p}p$) has
been called the odderon \cite{odd}; phenomenological fits to soft data
seem to indicate that its contribution to soft interactions at high
energies is negligeable, although this is still a matter of debate \cite{disc}.

There have been several attempts to relate Regge theory to QCD. Within the
framework of perturbative QCD, in the limit $\alpha_s \ln{s} \sim 1$
(leading-log approximation in $s$), the so-called
Balitsky-Fadin-Kuraev-Lipatov \cite{bfkl} (BFKL)
pomeron appears as a fixed cut in the $j$-plane with intercept
$1+\omega_{BFKL}(0)=1+(N_c \alpha_s/\pi) 4\ln{2}$ ($\simeq 1.5$ for $N_c=3$
and $\alpha_s\simeq 0.2$); it corresponds to the bound state of two $t$-channel
reggeized gluons. Under the same approximation the odderon appears as
the bound state of three gluons in a symmetric colour configuration,
given by the solution of the BKP \cite{bkp} equation.

Attempts to solve the odderon problem have gone in different directions.
On the one hand two-dimensional conformal techniques have been
applied \cite{integrable,kor}. On the other hand variational
methods have been
used, both with conformal invariant \cite{gln1,gln2} and
polynomial \cite{us} trial functions. This constitutes a first step
towards the unitarization \cite{unit} of QCD at high energies, which,
together with next-to-leading-log corrections \cite{nll}, are expected to
give full consistency to the whole approach \cite{muecia}.

In this contribution we will treat the following aspects \cite{us}: In Sect.
\ref{sec:arg} arguments will be presented for the odderon intercept
being exactly unity. In Sect. \ref{sec:var} a variational method will be
proposed. Finally in Sect. \ref{sec:res} some numerical results
and comparison to other calculations will be shown.

\section{Argument for intercept equal to unity}\label{sec:arg}

The BKP equation for three gluons with transverse momenta $q_1,q_2,q_3$
can be written in a Hamiltonian form with $E=1-j$:
\be
H\psi=E\psi,\ \ H=T_{1}+T_{2}+T_{3}+U_{12}+U_{23}+U_{31}.\label{eq:3}
\ee
In units of $N_c\alpha_s/\pi$,
\be
T_{1}=-\omega(q_{1})=\frac{\eta(q_{1})}{4\pi}
\int \frac{ d^{2}q'_{1}}{\eta(q'_{1})\eta(q_{1}-q'_{1})}\ \ ,\label{eq:4}
\ee
with
$\eta(q)=q^2+m^2$, is the gluon Regge trajectory. In (\ref{eq:3})
the dependence on $m$
vanishes, so the BKP equation is infrared stable \cite{bkp,jar}.

The $U_{ik}$'s are integral operators which act with measure
\bea
&d\mu=d^{2}q_{1}d^{2}q_{2}d^{2}q_{3}\delta^{(2)}
(q_{1}+q_{2}+q_{3}-q)/
[\eta(q_{1})\eta(q_{2})\eta(q_{3})],&\nonumber\\
&U_{12}(q_{1},q_{2},q_{3}|q'_{1},q'_{2},q'_{3})=
\eta(q_{3})\delta^{(2)}(q_{3}-q'_{3})
V_{12}(q_{1},q_{2}|q'_{1},q'_{2}).&\label{eq:5}
\eea
Here the
$V_{ij}$'s are BFKL interaction kernels for
2 gluons in a vector colour state (a factor
1/2 with respect to the vacuum channel appears):
\be
V_{12}(q_{1},q_{2}|q'_{1},q'_{2})=
-\frac{1}{4\pi}
\left [\frac{\eta_(q_{1})\eta_(q'_{2})+\eta_(q'_{1})\eta_(q_{2})}
{\eta(q_{1}-q'_{1})}
-\eta(q_{1}+q_{2})\right ]. \label{eq:6}
\ee

Due to the bootstrap identity \cite{bootstrap}:
\be
\int\frac{d^{2}q'_{1}}{\eta(q'_{1})\eta(q_1+q_2-q'_{1})}V_{12}
(q_{1},q_{2}|q'_{1},q'_{2})
=\omega(q_{1})+\omega(q_{2})-\omega(q_{1}+q_{2}),\label{eq:7}
\ee
for $q=q_1+q_2+q_3=0$, $\psi_{B}(q_{1},q_{2},q_{3})=\psi_{0}$ is a solution
with maximal symmetry which gives $E=0$.
This solution does not fullfil the gauge invariance requirement
$\psi_B(q_i=0)=0$ and for $m\rightarrow 0$
it
decouples from the physical spectrum, but still offers a lower bound for
the energy (as the state with
$|n|=1$, $\nu=0$ for the $|n|=1$ sector in the BFKL pomeron \cite{86}). More
elaborated mathematical arguments can be found in \cite{us}.

\section{Variational method}\label{sec:var}

Setting $m=0$, and redefining
$\psi\longrightarrow\prod_{i=1}^{3}q_{i}^{2}\psi$, we have
\be
H\psi=E\prod_{i=1}^{3}q_{i}^{2}\psi,\ \  H=(1/2)(H_{12}+H_{23}+H_{31})
\label{eq:8}
\ee
and, in a mixed representation ({\bf C} is Euler's constant),
\bea
H_{ik}\psi&=&\prod_{j=1}^{3}q_{j}^{2}(\ln q_{i}^{2}q_{k}^{2}+4{\bf
C})\psi
+\prod_{j=1,j\neq i,k}^{3}q_{j}^{2}\left [q_{i}^{2}\ln
(r_{ik}^{2}/4)\,q_{k}^{2} +(i\leftrightarrow
k)\right ]\psi \nonumber \\
&+&2(q_{i}+q_{k})^{2}\psi(r_{i}-r_k=0).\label{eq:9}
\eea

As usually, the variational approach provides us with an upper bound
for the ground state energy ($\equiv$ lower bound for the intercept)
and consists in finding a minimum of
the functional
\bea
&\Phi=\int\prod d^{2}q_{i}\psi^{\ast}H\psi,& \nonumber\\
&\int\prod d^{2}q_{i}\psi^{\ast}\prod_{j=1}^{3}q_{j}^{2}\psi=1.&\label{eq:10}
\eea
As trial function we choose a linear combination of one-gluon
functions:
\be
\psi(r_{1},r_{2},
r_{3})=\sum_{\alpha_{1},\alpha_{2},\alpha_{3}}
c_{\alpha_{1},\alpha_{2},\alpha_{3}}\prod
_{i=1}^{3}
\psi_{\alpha_{i}}(r_{i}),\label{eq:11}
\ee
with
\be
\int
d^{2}r\psi_{\alpha}^{\ast}q^{2}\psi_{\alpha'}=\delta_{\alpha,\alpha'},\
\ \sum_{\alpha_{1},\alpha_{2},\alpha_{3}}
|c_{\alpha_{1},\alpha_{2},\alpha_{3}}|^{2}=1\label{eq:12}
\ee
and $c_{\alpha_{1},\alpha_{2},\alpha_{3}}$ fully symmetric in
$\alpha_{1},\alpha_{2},\alpha_{3}$.

Finally all is reduced to diagonalize the matrix
${\cal
E}_{\alpha_{1},\alpha_{2},\alpha'_{1},\alpha'_{2}}$ defined by
\be
\Phi=\frac{3}{2}
\sum_{\alpha_{1},\alpha_{2},\alpha'_{1},\alpha'_{2},\alpha_{3}}
c^{\ast}_{\alpha_{1},\alpha_{2},\alpha_{3}}
c_{\alpha'_{1},\alpha'_{2},\alpha_{3}}{\cal
E}_{\alpha_{1},\alpha_{2},\alpha'_{1},\alpha'_{2}},\label{eq:13}
\ee
multiplied by the identity for the third gluon and symmetrized
in $(1,2,3)$ and $(1^\prime,2^\prime,3)$.

Our concrete choice of trial basis are the harmonic
oscilator eigenfunctions:
\bea
&\psi_{\alpha}({\bf r})=\psi_{k,l}(z)\exp il\phi,\ \ z=\ln{r^2},\ \
\psi_{k,l}(-\infty)=0,& \nonumber \\
&\xi_{k}(z)=c_{k}H_{k}(z)\exp
(-z^{2}/2)=(\partial+(1/2)|l|)\psi_{k,l}(z),& \label{eq:14}
\eea
with $H_k(z)$ the Hermite polynomials.

\section{Numerical results}\label{sec:res}

From our numerical experience the best results are obtained with
$|l|\leq l_{max}$, $k=0,1,...,(l_{max}+1=r)$. They
are shown in Table \ref{tab:1}.
\begin{table}[htb]
\caption{Lowest eigenvalues of
(\protect{\ref{eq:8}}),(\protect{\ref{eq:9}})
for the two- ($\epsilon_{2}$) and
three-gluon ($\epsilon_{3}$) bound states, using our trial functions
labelled by $r$. $-2.773\simeq -4\ln{2}$
is the exact value for $\pi\ \omega_{BFKL}(0)/(N_c\alpha_s)$.}
\vspace{0.4cm}
\begin{center}
\begin {tabular}{ccc}  \hline \hline
  $r$ & $\epsilon_{2}$ & $\epsilon_{3}$\\\hline
   1  &  0.968        & 0.968     \\
   2  &  0.022        & 0.605     \\
   3  &$-0.475$       & 0.454     \\
   4  &$-0.743$       & 0.379     \\
   5  &$-0.912$       & 0.331     \\
   6  &$-1.032$       & 0.298     \\
      &               &            \\
$\infty$& $-2.773$    &            \\\hline \hline
\end{tabular}
\end{center}
\label{tab:1}
\end{table}

These results are related to the corresponding intercepts as
\be
1+\omega_{BFKL}(0)= 1-\frac{N_c\alpha_s}{\pi} \epsilon_2,\ \
1+\omega_{odd}(0)= 1-\frac{N_c\alpha_s}{\pi} \frac{3}{2} \epsilon_{3}.
\label{eq:15}
\ee
The dimension of the matrix grows, for the three-gluon case, from 12
($r=2$) to 3368 ($r=6$). For $r=1$ the potential energy $U_{ik}$
vanishes and one
gets the kinetic energy per gluon, i.e. $\epsilon_2= \epsilon_3$ in this
case. It can be seen that the convergence
of the method is quite slow.

In Table \ref{tab:2} we compare our results to other calculations. Our
lower bound is
compatible with our previous argument of $\epsilon_3=0$ but weaker than
that of \cite{gln2}, which gives a supercritical odderon; this
discrepancy may have a numerical origin.
As to the results of  \cite{kor},
they are in principle exact but contain
some semiclassical approximation whose
reliability is difficult for us to quantify. So our conclusion is that
the odderon intercept
lies in the interval $0.91\div 1.79$ (for $N_c=3$ and
$\alpha_s=0.2$).
\begin{table}[htb]
\caption{Comparison of our results to other calculations of the
odderon intercept.}
\vspace{0.4cm}
\begin{center}
\begin {tabular}{ccc}  \hline \hline
& $-(3/2)\epsilon_3$ & $1+\omega_{odd}(0)$ for \\
& & $N_c=3$, $\alpha_s=0.2$
\\ \hline
Our result & $\geq -0.45$ & $\geq 0.91$ \\
Ref. \protect{\cite{gln2}} & $\geq 0.37$ & $\geq 1.07$ \\
Ref. \protect{\cite{kor}}   & $2.41$ & 1.46 \\
Upper bound from \protect{\cite{gln1}}
& $\leq 4.16$ & $\leq 1.79$ \\ \hline\hline
\end{tabular}
\end{center}
\label{tab:2}
\end{table}

To quantify the existing uncertainty in more practical terms, let us
consider the process $\gamma \mbox{p}\rightarrow \eta_c\mbox{p}$, i.e.
diffractive photoproduction of $\eta_c$. The corresponding cross-section
has been estimated \cite{eta} to be
$\sigma(\gamma \mbox{p}\rightarrow \eta_c\mbox{p}) = D \times (47\div 100)
\ \ \mbox{pb}, \ \ D=(\overline{x})^{-2\omega_{odd}(0)}$.
From the previous considerations, $D$ (for $N_c=3$ and
$\alpha_s=0.2$) lies in the range $0.3\div 55000$
for HERA ($\overline{x}\simeq 10^{-3}$). Clearly this reaction, in case it
could be studied experimentally, is very sensitive to the value of the
odderon intercept and offers a good opportunity to measure it.

In conclusion we have studied the odderon singularity as solution of
the
BKP equation for three colour-symmetric gluons
and given an argument for the odderon intercept to
be equal to unity. Besides a variational method to compute it has been
presented and its results compared to other calculations. The odderon
intercept lies in the interval $1-(N_c \alpha_s/\pi)0.45\leq
j_{odd}(0)=1+\omega_{odd}(0) \leq 1+(N_c \alpha_s/\pi)4.16$.

\section*{Acknowledgments}
The authors thank J. Bartels,
P. Gauron, G. P. Korchemsky, L. N. Lipatov, B. Nicolescu
and S. Wallon for useful discussions and the organizers for such a nice
meeting. N. A. also thanks Direcci\'on General de Investigaci\'on
Cient\'{\i}fica y T\'ecnica of Spain for
financial support.

\end{document}